\documentclass{article}

\usepackage{xcolor}
\definecolor{mygreen}{RGB}{0, 176, 80}
\definecolor{myblue}{RGB}{0, 112, 192}
\definecolor{mypurple}{RGB}{112, 48, 160}

\usepackage{arxiv}

\usepackage[utf8]{inputenc} 
\usepackage[T1]{fontenc}    
\usepackage{hyperref}       
\usepackage{url}            
\usepackage{booktabs}       
\usepackage{amsfonts}       
\usepackage{nicefrac}       
\usepackage{microtype}      
\usepackage{lipsum}
\usepackage{graphicx}
\usepackage{authblk}
\usepackage{textcomp, gensymb}
\usepackage{float}
\graphicspath{ {./images/} }
\usepackage{mathtools}
\usepackage{amsmath,amssymb}
\usepackage{comment}
\usepackage{mathrsfs}
\usepackage{graphicx,nicefrac}
\usepackage{float}
\usepackage{amssymb}

\usepackage{comment}
\usepackage{scalerel}
\usepackage[overload]{empheq}
\usepackage{textgreek}

\usepackage{algorithm}
\usepackage{algpseudocode}
\usepackage{enumitem}

\usepackage{mathrsfs}
\usepackage{graphicx,nicefrac}

\usepackage[utf8]{inputenc}
\usepackage[english]{babel}

\usepackage[margin=1cm]{caption}
\usepackage{comment}
\usepackage{scalerel}
\usepackage[overload]{empheq}

\usepackage{float}
\usepackage{siunitx}

\usepackage{mathtools}

\usepackage{authblk}
\title{Rapid 3D imaging at cellular resolution for digital cytopathology with a multi-camera array scanner (MCAS)}

\author[1]{Kanghyun Kim}
\author[1]{Amey Chaware}
\author[1]{Clare B. Cook}
\author[1]{Shiqi Xu}
\author[2]{Monica Abdelmalak}
\author[3]{Colin Cooke}
\author[1,4]{Kevin C. Zhou}
\author[5]{Mark Harfouche}
\author[5]{Paul Reamey}
\author[5]{Veton Saliu}
\author[5]{Jed Doman}
\author[5]{Clay Dugo}
\author[5]{Gregor Horstmeyer}
\author[2]{Richard Davis}
\author[2]{Ian Taylor-Cho}
\author[2]{Wen-Chi Foo}
\author[1]{Lucas Kreiss}
\author[2]{Xiaoyin Sara Jiang}
\author[1,3,5,*]{Roarke Horstmeyer}

\affil[1]{Department of Biomedical Engineering, Duke University, Durham, NC 27708, USA}
\affil[2]{Department of Pathology, Duke University Medical Center, Durham, NC 27708, USA}
\affil[3]{Department of Electrical and Computer Engineering, Duke University, Durham, NC 27708, USA}
\affil[4]{Department of Electrical Engineering and Computer Science, University of California, Berkeley, Berkeley, CA 94720, USA}
\affil[5]{Ramona Optics Inc., 1000 W Main St., Durham, NC 27701, USA}

\affil[*]{Corresponding author: roarke.w.horstmeyer@duke.edu}

\begin{document}
\maketitle
\begin{abstract}
Optical microscopy has long been the standard method for diagnosis in cytopathology. Whole slide scanners can image and digitize large sample areas automatically, but are slow, expensive and therefore not widely available. Clinical diagnosis of cytology specimens is especially challenging since these samples are both spread over large areas and thick, which requires 3D capture. Here, we introduce a new parallelized microscope for scanning thick specimens across extremely wide fields-of-view (54$\times$72 mm$^2$) at 1.2 and 0.6 \SI{}{\micro\metre} resolutions, accompanied by machine learning software to rapidly assess these 16 gigapixel scans. This Multi-Camera Array Scanner (MCAS) comprises 48 micro-cameras closely arranged to simultaneously image different areas. By capturing 624 megapixels per snapshot, the MCAS is significantly faster than most conventional whole slide scanners. We used this system to digitize entire cytology samples (scanning three entire slides in 3D in just several minutes) and demonstrate two machine learning techniques to assist pathologists: first, an adenocarcinoma detection model in lung specimens (0.73 recall); second, a slide-level classification model of lung smears (0.969 AUC).
\end{abstract}

\section{Introduction}
Optical microscopy has long suffered from a trade-off between field-of-view (FOV) and imaging resolution. Concretely, a single-objective optical microscope is typically limited in its etendue or spatial bandwidth product, which defines the maximum amount of information it can capture~\cite{lohmann1989scaling, brady2012multiscale, mcconnell2016novel}.

\begin{figure*}[ht]
    \centering
    \includegraphics[scale = 0.17]{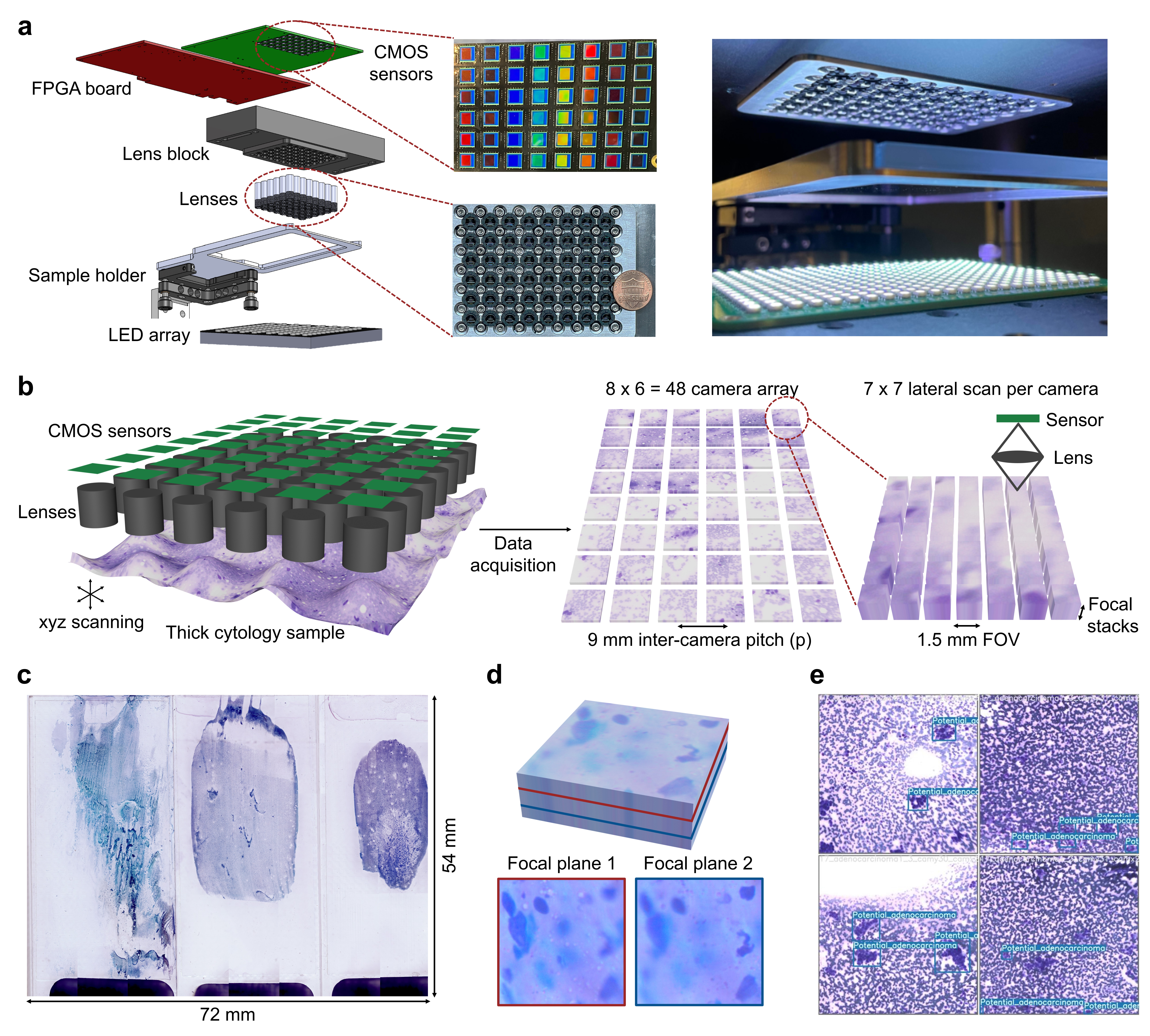}
    \caption{\textbf{Architecture of the Multi-Camera Array Scanner (MCAS).} \textbf{a} The MCAS with 6$\times$8 sensor array (48$\times$ 13-megapixel CMOS sensors and associated objective lenses). \textbf{b} Rapid 3D imaging with 48 compact microscopes, each providing a distinct 1.5 mm field of view. 7$\times$7 lateral scanning grid addresses the 9 mm inter-camera gaps and axial scanning captures surface irregularities of thick specimens. (\textbf{c}) Simultaneous digitization of 3 thick cytology smears with MCAS produces high-resolution stitched image from 2,352 unique captures. (\textbf{d}) High-speed focus adjustment capabilities enable 3D slide examination. (\textbf{e}) The MCAS can facilitate efficient cytopathology consultations with rapid 3D whole slide imaging and integrating machine learning capabilities.}
    \label{fig:teaser}
\end{figure*}

This presents a challenge in many applications of optical microscopy. A key example is the examination of cytopathological specimens, which requires imaging a large area (cytology smears often cover up to 75 mm$\times$25 mm) at high resolutions ($\sim$\SI{1}{\micro\metre}), as the features of interest are at a cellular or sub-cellular level. A commonly performed cytopathological assay is fine needle aspiration (FNA) cytology. It is a widely used first-line modality to diagnose cancers of various sites including thyroid, lung, and pancreas~\cite{donnelly2013optimal, idowu2010lung}. In this procedure, cell samples are drawn from the site with a thin needle, smeared on a glass slide, and stained. A pathologist then visually inspects the smears to localize the cancer cells. Visual inspection of FNA cytology smears is the gold standard for diagnosing a large variety of cancers in these specimens ~\cite{poostchi2018image, cibas2013cytology}. 

In conventional cytopathology, objective lenses used for inspecting FNA smears range from 10X to 40X, with magnification values depending on prioritizing throughput (large FOV, low magnification) or resolution (small FOV, high magnification). The FOV of a conventional microscope is 0.5-2 mm in diameter, with a depth of field (DOF) on the order of \SI{1}{\micro\metre}. Most FNA smears extend over several square centimeters area and have variable thicknesses, that can be over $\sim$\SI{50}{\micro\metre} thick, with cellular matter distributed at a variable density across the specimen volume. This requires a human observer to manually move the sample in all three axes for inspection, which can be time-consuming, error prone, and resource-intensive, particularly in cases with multiple slides, delaying diagnosis. While all pathology applications are time-sensitive, a quick response is critical for rapid on-site evaluation (ROSE), where slides are evaluated in real-time during a procedure. ROSE enables immediate triage and ensures sample adequacy, minimizing the need for repeated sampling. However, ROSE is resource-intensive and faces challenges, especially due to staffing shortages common in many healthcare settings~\cite{garcia2015american}.

One potential avenue to enhance pathology workflow accuracy and efficiency is digitizing pathology slides using a whole slide imaging (WSI) system~\cite{farahani2015whole}, which mechanically scan slides to create gigapixel composites for remote viewing and diagnosis. WSI also allows for the use of software tools for image analysis, such as segmentation or identifying cancerous cells~\cite{khened2021generalized}. While this technology is widely utilized for imaging thin tissue sections~\cite{aeffner2019introduction}, there are few effective and rapid solutions for digitizing and thus using software to examine thick cytology smears~\cite{eccher2020current}.

Given that the thick cytology smears necessitate WSI systems to scan significantly larger slide areas both laterally and axially (i.e., in 3D), the volume of image data can increase by several orders of magnitude compared to thin pathology samples. Consequently, cytology slide scanning times can often take over an hour per slide~\cite{donnelly2013optimal, wright2013digital, fan2016method}, potentially prolonging cytopathological assessments~\cite{michael2020rapid}. Therefore, conventional WSI is of limited diagnostic utility for cytology smears and is currently predominantly used within research settings. Owing to the scarcity of WSI for thick cytology smears, there is a paucity of software tools developed for automated analysis of these smears~\cite{teramoto2017automated}. Therefore, the absence of appropriate WSI technology precludes the possibilities for remote viewing and virtual diagnosis of thick cytology smears.

Several alternative methods for rapidly imaging large areas have been proposed, such as Structured Illumination Microscopy (SIM)~\cite{gustafsson2000surpassing} and Fourier Ptychography~\cite{konda2020fourier, zheng2014fourier, zhang2024fpm, chen2022rapid}. These techniques manipulate the microscope's illumination instead of mechanically scanning the sample, acquiring time-sequential images that are computationally merged to create large, high-resolution composites. While these methods significantly improve resolution and allow for the use of low-magnification, large-FOV objective lenses, their speed is hindered by the time required to capture and process multiple variably illuminated images, in particular for significantly thicker specimen ($>$ \SI{50}{\micro\metre}). Furthermore, the FOV of these technologies is still considerably smaller than a typical cytology slide (covering up to 75 x 25 mm).

While a single-lens microscope can only capture limited information, a design that uses multiple units of lenses and image sensors can parallelize the imaging procedure and thus capture much more information per snapshot. Such multi-aperture systems were initially developed to image distant macroscopic objects~\cite{brady2012multiscale}. Array-based imaging systems later were used to capture light fields~\cite{wilburn2005high}, make compact optical designs~\cite{tanida2001thin}, enhance close-up photography~\cite{marks2011close}, and more~\cite{cribb2015high,chan2019parallel,cibir2023complexeye}. In our previous work, we developed an array-based system called the multi-camera array microscope (MCAM) to translate this concept to the microscopic world and image large areas ($\approx$100 cm$^2$) at a relatively high resolution (\SI[parse-numbers=false]{5-20}{\micro\metre})~\cite{harfouche2023imaging,yang2023multi}. This combination of FOV and resolution makes the MCAM ideal for imaging freely moving model organisms such as zebrafish, harvester ants, and \textit{Drosophilia} flies~\cite{zhou2023parallelized}. However, the previous MCAM design had limited axial scanning capabilities, offered resolutions insufficient to resolve details at the cellular level, and was not outfitted with parallelized scanning technology required to rapidly and fully digitize thick cytopathology specimens.

Digitizing thick cytology smears is challenging due to the need for high resolutions, the 3D nature of samples, and extremely large resulting datasets, which can be on the order of 100 GB. This has led to a lack of effective solutions for quickly imaging entire smears, hindering cytopathologists from fully utilizing digital pathology benefits. To overcome these challenges and rapidly image thick microscopic samples at high resolution across a large field of view, we introduce a multi-camera array \textit{scanner} (MCAS). The MCAS comprises 48 individual image sensors organized into a tightly packed (9 mm spacing) 6$\times$8 grid (Fig.~\ref{fig:teaser}a). Each camera is outfitted with a custom-designed, high-resolution objective lens capable of resolving sub-cellular features. Additionally, the MCAS is equipped with sample stages and associated novel firmware to rapidly translate up to 3 slides at once to complete 3D scanning. We showcase the MCAS digitizing full, axially thick smears into multi-gigapixel composites at two unique resolutions - \SI{1.2}{\micro\metre} and \SI{0.6}{\micro\metre} full-pitch resolution (comparable with standard 10X and 20X objective lenses) - by scanning across depths up to \SI{150}{\micro\metre}. This system can capture the entire volume of a thick sample in a parallelized fashion, theoretically achieving a 48-fold increase in imaging speed (Fig.~\ref{fig:teaser}b), compared to a single microscope of the same magnification. With this significant speed increase, we demonstrate applications of our new microscope for three potentially new cytopathology workflows: 1) for remote 3D whole-slide viewing via a custom-designed interface, 2) for remote assessment of thyroid FNA slide adequacy, and 3) to detect and predict the presence of adenocarcinoma, the most common form of lung cancer (Fig.~\ref{fig:teaser}c-e).

\section{Results}
\subsection{Multi-camera array scanner design}
The MCAS system contains an array of 48 unique 13-megapixel (MP) CMOS image sensors (Onsemi AR1335, 4208H$\times$3120V, $\delta$ = \SI{1.1}{\micro\metre}), each equipped with its own identical custom-designed objective lens. All 48 sensors are arranged in a compact, uniform grid on a single circuit board. Each lens captures data from a distinct, non-contiguous sample area, as depicted in Fig.~\ref{fig:teaser}b. We used two different custom-designed finite-conjugate objective lenses, sourced from Edmund Optics and Avantier Inc. to compare their resolution and understand the trade-off between resolution, per-camera FOV, and total 3D scan time. The lenses from Edmund Optics feature 2.2X magnification, a 0.3 numerical aperture (NA), and a 3.05 mm working distance, while the lenses from Avantier Inc. offer 6.89X magnification, a 0.5 NA, and a 1.74 mm working distance, corresponding to conventional 10X and 20X lenses respectively. We achieved \SI{1.2}{\micro\metre} full-pitch resolution for the 0.3 NA lens and \SI{0.6}{\micro\metre} full-pitch resolution for the 0.5 NA lens. This significantly improves upon previous efforts in multi-camera array designs to facilitate operation in the sub-cellular resolution regime~\cite{harfouche2023imaging}.
 
For whole-slide 3D imaging of FNA smears, we aimed to capture data over a 54$\times$72 mm$^2$ FOV, which fits three microscope slides. The number of scan positions needed to fill FOV gaps between adjacent micro-cameras at a given magnification $M$ increases with inter-camera pitch $p$ ~\cite{harfouche2023imaging}. Therefore, we designed our new imaging array to have as small a pitch ($p=9$ mm) as currently possible given the size, packaging, and readout electronic constraints of our CMOS sensors to increase the speed of per-slide capture. Unlike standard microscopes or WSI scanners, which require stage translation across 75 mm to capture the whole slide, the MCAS only needs to scan as far at the inter-camera distance $p$ (9 mm). We used 3-axis motorized stages from Zaber Technologies for precise specimen movement across 9 mm laterally and to axially refocus. For our 0.3 NA lens design, 7$\times$7 lateral scanning per axial slice was employed, while 20$\times$20 lateral scanning was used for the 0.5 NA design. To fully capture 3D FNA smears, lateral scanning is repeated across multiple axial slices, with an axial step and total distance that can be user-selected (see details below). At each scan location, all 48 micro-cameras synchronously (within \SI{6}{\micro\metre} max. offset) capture 0.63 GP (48$\times$4208$\times$3120) per snapshot. Data is transferred through a field-programmable gate array (FPGA) at up to approximately 5 GP/second (or 5 GB/second), which allows the system to operate at up to 8 synchronized frames per second. The frame rate can be increased by pixel binning or reducing the sensors' active area.

\subsection{Rapid whole-slide imaging of lung specimens}
\begin{figure*}[ht]
    \centering
    \includegraphics[scale = 0.18]{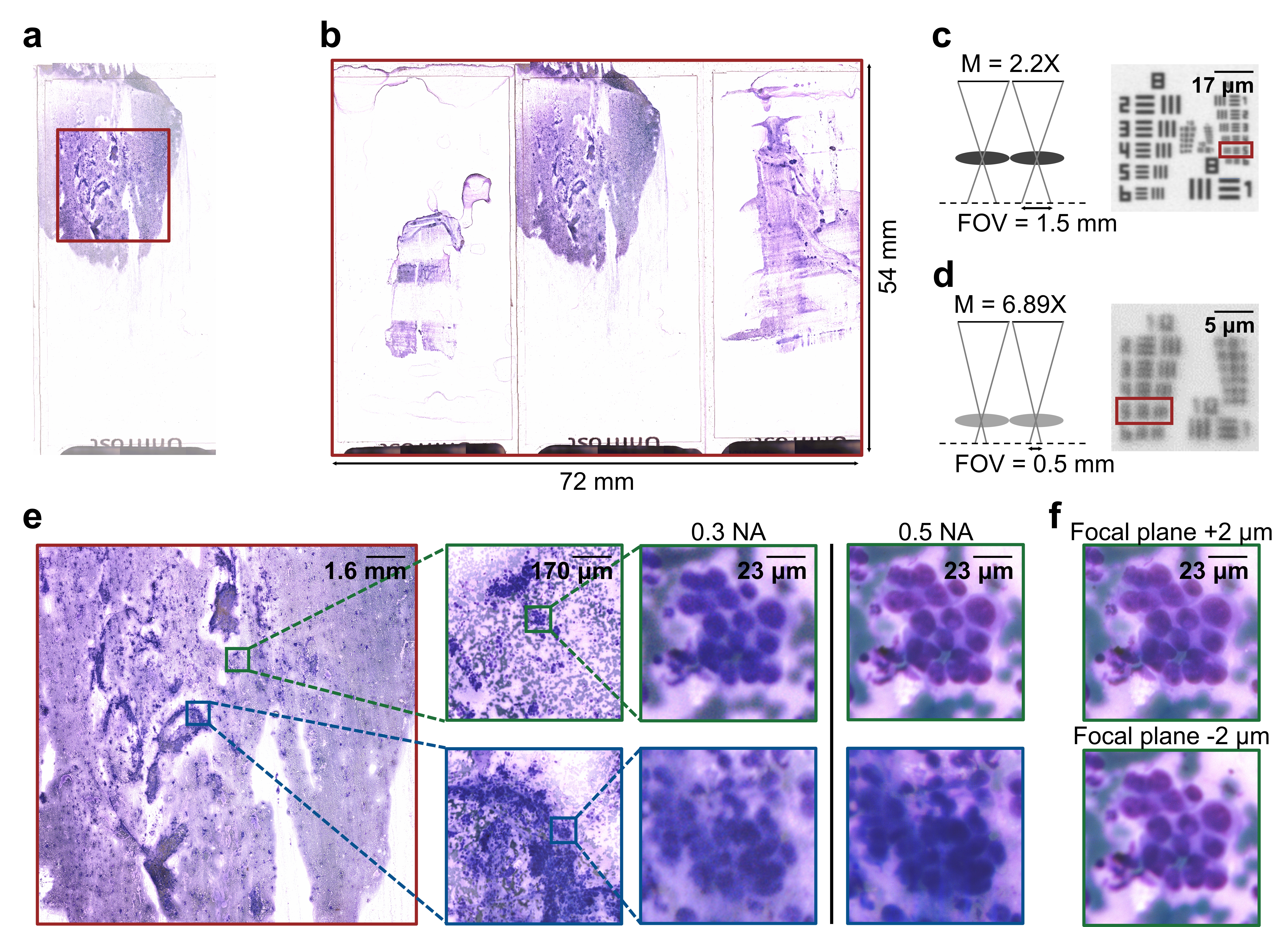}
    \caption{\textbf{FOV and resolution demonstration of the MCAS.} \textbf{a}-\textbf{b} FOV comparison between a standard whole slide imaging and the MCAS. Each red box represents the FOV from each system. (\textbf{a}) Standard whole slide imaging captures a focal stack of 15$\times$15 mm$^2$ area in 30-60 minutes using 0.5 - 0.8 NA lenses. (\textbf{b}) The MCAS captures three whole slides, covering a 3D stack of 54$\times$72 mm$^2$ area, in 5 and 40 minutes with 0.3 and 0.5 NA lenses, respectively. \textbf{c}-\textbf{d} Resolution and FOV comparison between our custom-designed 0.3 NA and 0.5 NA lenses (full-pitch resolutions of \SI{1.2}{\micro\metre} and \SI{0.6}{\micro\metre}, respectively). \textbf{e} MCAS image comparison between the two custom-designed lenses. The 0.5 NA lens provides higher resolution, although at the cost of increased data acquisition time. \textbf{f} Demonstration of focus adjustment capability. Each image is taken from \SI{2}{\micro\metre} above and below the central focal plane in (\textbf{e}).}
    \label{fig:resolution_comparison}
\end{figure*}

We demonstrate rapid whole-slide 3D imaging capabilities with lung FNA specimens. Specifically, we scanned 16 unique Diff-Quik stained cytology samples from archival cases, including adenocarcinoma-positive and benign samples. The samples generally extended across an axial range of \SI{120}{\micro\metre}, defining our axial scan distance. Unlike standard slide scanners which are limited to capturing a small portion of a slide ($<$ 2 cm$^2$), the MCAS can image the entire volume of three slides simultaneously. Using 0.3 NA lenses and an axial step size of \SI{5}{\micro\metre}, we achieved imaging in less than 5 minutes for all 3 slides (54$\times$72 mm$^2$ area), equating to less than 2 minutes per slide. In comparison, today's WSI scanners, which are used for small area (15×15 mm²) scanning, typically require 1 minute for 2D and tens of minutes for 3D capture. Our MCAS makes this possible at a fraction of the cost of standard WSI systems. Images acquired with the 0.5 NA lens display enhanced resolution which facilitates improved discernment of adenocarcinoma-positive regions, as highlighted in Fig.~\ref{fig:resolution_comparison}e. However, this increase in the information captured comes at the cost of a longer acquisition time. The 0.5 NA lens requires a finer axial step size (\SI{1}{\micro\metre}) for effective focal stack capture due to its shallower depth of field. This increases total acquisition from five minutes when imaging with the 0.3 NA lenses (forty focal stacks, 7$\times$7 scans per slice) to forty minutes with the 0.5 NA lenses (forty focal stacks, 20$\times$20 scans per slice) (Fig.~\ref{fig:workflow}b).

The 3D raw data scanned by the MCAS goes through multiple processing steps before it is displayed to the pathologists. First, the most in-focus planes within the z-stack are identified using a contrast metric across all lateral scanning positions. The most in focus images are then stitched using customized code that is based on open-source software ~\cite{hugin} to create an all-in-focus view. To facilitate focus adjustment during 3D slide examination, we next stitch adjacent planes around the in-focus plane. Navigation through the 16 GP stitched images per plane (80 GP total for 5 axial planes) is facilitated by our custom viewing software - Gigaviewer (\url{https://gigazoom.rc.duke.edu}). Gigaviewer provides a navigation experience similar to various popular applications such as Google Maps, allowing continuous zoom across multiple scales combined with panning. It also provides the ability to axially refocus imagery at any location/zoom setting for a complete remotely viewing experience for 3D digitized specimens. Further details on these processes are available in the Methods section.

\subsection{Adequacy assessment with thyroid specimens}
The MCAS's rapid 3D data acquisition and Gigaviewer navigation enables remote inspection of cytology smears. This capability can allow the MCAS to improve cytopathology workflows. To showcase this, we first apply the MCAS to aid with the task of smear adequacy assessment. 

As discussed above, FNA cytology is widely used to diagnose lesions of various sites by drawing cellular material using a fine needle, smearing it onto a glass slide, and staining it for microscopic evaluation. A pathologist later examines one or more smears for diagnostic assessment. It thus is important to ensure that each smear is adequately prepared (i.e., contains sufficient cellular material for effective diagnosis) during FNA procedure. To assess the adequacy of the smear, rapid on-site evaluation (ROSE) of FNA smears is carried out in real-time by a pathologist or cytotechnologist. If the smear is determined to be inadequate, then sample extraction can be repeated immediately, saving both patients and medical practitioners the time and cost associated with repeated procedures. ROSE is common, with more than 6000 procedures performed per year at many hospitals~\cite{lin2019rapid}. 

\begin{figure}[ht]
    \centering
    \includegraphics[scale = 0.2]{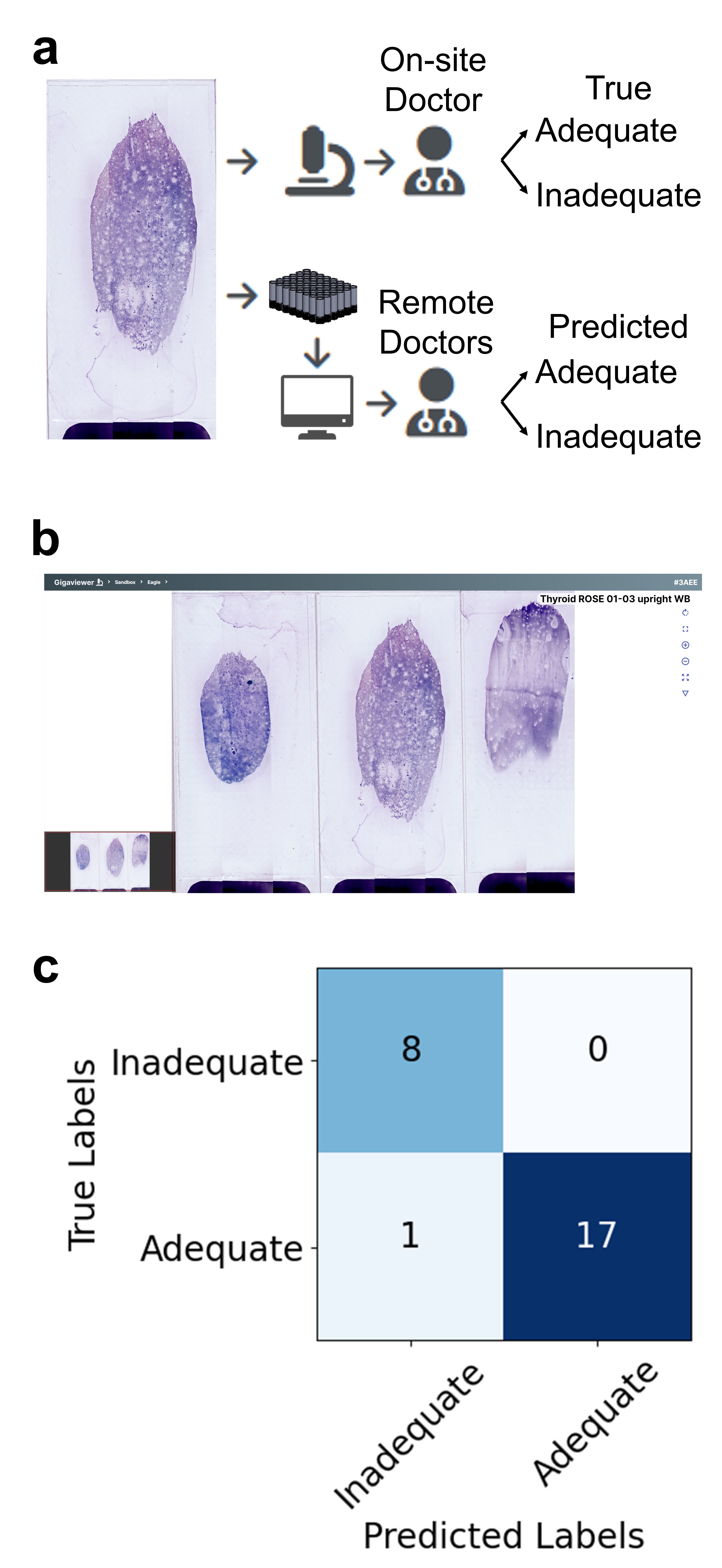}
    \caption{\textbf{Evaluation of thyroid specimen adequacy.} \textbf{a} One pathologist conducted the adequacy assessment using a standard microscope (True Labels). Three more pathologists used the MCAS and Gigaviewer to remotely conduct the assessment (Predicted Labels). \textbf{b} Gigaviewer interface, which is a customized viewer for remote viewing (Supplementary Movie 1). \textbf{c} Assessments for each slide from the three pathologists are combined using majority vote, and the results are then compared to evaluate the consistency between both methods.}
    \label{fig:rose_assessment}
\end{figure}

Typically, ROSE is performed in dedicated sites, where a pathologist observes the smear under a microscope to decide smear adequacy. The rapid digitization and the remote viewing capability offered by MCAS can eliminate the need for slide transportation, the physical presence of the pathologist, and can provide an immediate digital record of each slide to make ROSE procedures faster and more streamlined.

As a first step towards such a demonstration, we tested the MCAS with remote ROSE on 26 Diff-Quik stained thyroid FNA smears (obtained from archival cases, anonymized before the experiment). First, one pathologist evaluated the smears for adequacy using a conventional microscope (the standard method for performing ROSE). 18 smears (69.2\%) were determined to be adequate, and 8 smears (30.8\%) were deemed inadequate. We treated these adequacy labels as being the ``True" labels. Then, we digitized all of the smears using the MCAS and displayed them to three other pathologists using the Gigaviewer interface. Each pathologist inspected all 26 smears digitally and classified them as adequate or inadequate (Fig.~\ref{fig:rose_assessment}). Adequacy labels were then assigned via a majority vote and we treated these as the ``Predicted" labels. Detailed results for each pathologist are available in Supplementary Fig. 1. We found that MCAS-based ROSE adequacy decisions were on par with the current standard and exhibited a sensitivity of 100\% and a specificity of 94.4\%. This evaluation highlights the MCAS's potential to optimize cytology workflows and improve efficiency in clinical settings.

\subsection{Suspected adenocarcinoma detection from lung specimens}
The volume of data produced by digitizing entire cytology smears in 3D is significant. To make the examination and analysis of this data more efficient, we next examined how machine learning algorithms could assist with detecting key regions of interest within MCAS image data. Concretely, we considered the diagnosis of lung cancer, which is frequently performed using FNA cytology. Lung cancer has the third highest incidence rate out of all cancers in the U.S., with over 230,000 new cases annually. It is also the leading cause of cancer death, representing 21\% of all cancer fatalities in the U.S. in 2023~\cite{siegel2023cancer}. We focus specifically on locating areas within the lung FNA smears that could potentially indicate the presence of adenocarcinoma of the lung, the most common type of lung cancer.

\begin{figure}[ht]
    \centering
    \includegraphics[scale = 0.2]{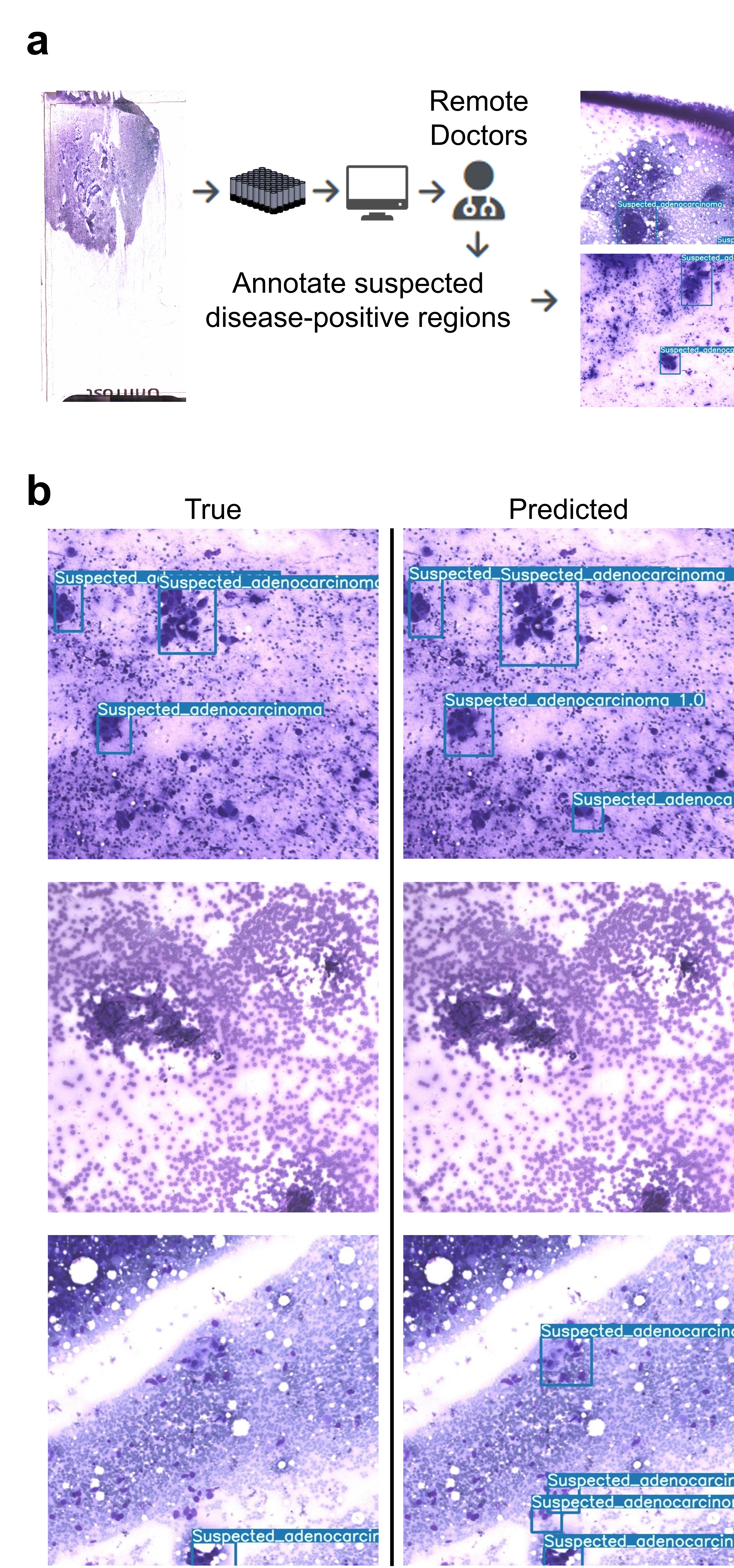}
    \caption{\textbf{Detection of suspected adenocarcinoma in lung specimens using YOLOv7.} \textbf{a} Pathologists annotate cancerous regions on MCAS images. \textbf{b} Machine learning results locate areas of suspected adenocarcinoma, marked as blue boxes, using a supervised learning algorithm. The trained model achieves a recall of 0.73.}
    \label{fig:object_detection}
\end{figure}

We used the MCAS (0.3 NA lenses) to image five unique FNA smears from patients diagnosed with either adenocarcinoma or benign conditions. For machine learning tasks, we did not stitch the data captured by the individual sensors into a composite. Each sensor of the MCAS captures a 3072$\times$3072 pixel image, which we then subdivided into four 1536$\times$1536 pixel patches for training. We presented these patches to clinicians, who annotated them by drawing bounding boxes around suspected adenocarcinoma-containing regions. A total of 1,794 patches were annotated and split into training, validation, and test sets, comprising 1201, 384, and 209 patches respectively. As stratification, all patches from any given sensor image were assigned to only one data split. We used this labeled data to train well-studied object detection algorithms to locate the suspected adenocarcinoma-positive regions in the smear images.

Convolutional neural network (CNN) based object detection has been utilized for the automated detection and diagnosis of various diseases, such as malaria in blood smears~\cite{kim2020multi}, tuberculosis (TB) in sputum smears~\cite{quinn2016deep}, and aiding cancer diagnoses from cytology smears~\cite{landau2019artificial}. Among the many neural network frameworks for object detection, we selected YOLOv7~\cite{wang2023yolov7} for its proven speed and robustness across multiple studies. The results of object detection are showcased in Fig.~\ref{fig:object_detection}, with the network's performance denoted by a mean average precision of 0.645 and a recall of 0.73. Our object detection results demonstrate the potential of rapid digitization with the MCAS for analyzing cytology smears through machine learning, thus offering additional avenues to reduce diagnosis time. A much larger dataset containing a diverse set of slides, ideally annotated by multiple clinicians, will facilitate a more comprehensive assessment of automated adenocarcinoma localization within 3D cytology smears.

\begin{figure*}[ht]
    \centering
    \includegraphics[scale = 0.27]{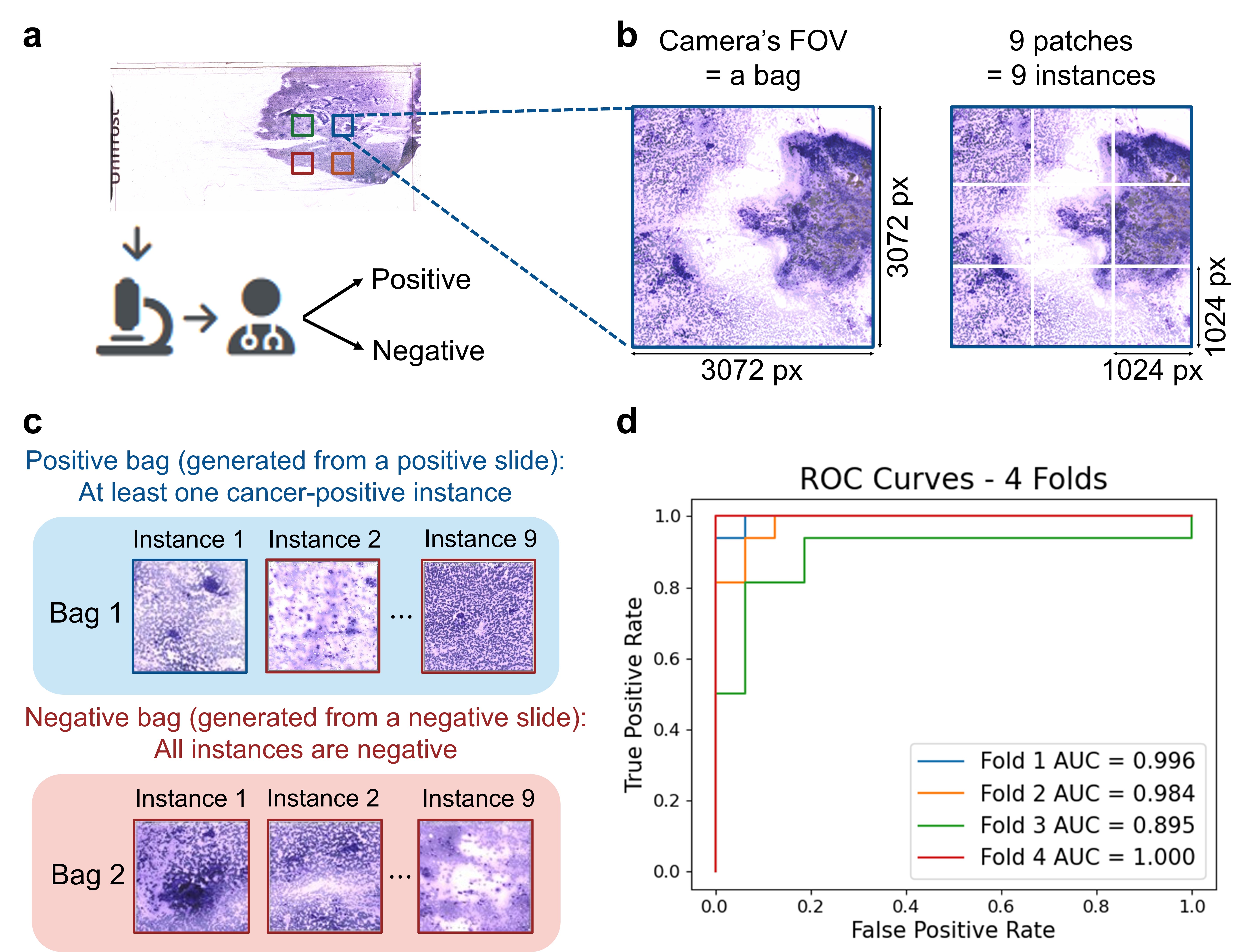}
    \caption{\textbf{Implementation of multiple instance learning (MIL) for accurate slide-level automated diagnosis.} \textbf{a} Clinicians assigns each slide a category at the slide level using standard microscope evaluation. \textbf{b}-\textbf{c} The MCAS scans slides, dividing each unique camera's FOV into 9 sub-patches to generate a `bag' dataset. \textbf{d} Performance of MIL slide-level classification (cancer-positive vs. negative) with four-fold cross-validation.}
    \label{fig:slide_level_classification}
\end{figure*}

\subsection{Slide-level adenocarcinoma classification from lung specimens}
In diagnosing lung cancer from FNA smears, pinpointing the spatial location of areas indicative of disease is not always essential~\cite{chong2023stepwise}. Instead, the primary objective is to classify the type of cancer, when detected, into one of several possible categories. This fact, combined with the challenges associated with obtaining accurate human cell-level annotations, led us to consider designing a slide-level classification network for cancer types. In this method, \emph{entire slides} are assigned a class label depending on the type of cancer they display (or a negative label if they are benign). Then, after imaging the slides, a machine learning algorithm is trained to classify each \emph{slide} into one of the cancer categories. For this work, we limit the scope to rapidly identifying whether a given slide contains adenocarcinoma or is benign via rapid MCAS imaging.

In a slide-level classification task, it is not guaranteed that every image patch that came from a given slide will contain features that tie it to its labeled class. Therefore, to perform slide-level classification, we need to employ a model capable of simultaneously assessing multiple image patches from a single slide to make an informed decision. This is possible using a technique known as Multiple Instance Learning (MIL) ~\cite{cooke2022multiple}. In MIL, the model is trained using labeled bags of instances rather than individual instances, where each bag contains many individual instances. For our binary classification task, it is assumed that if a bag is labeled positive, then at least one of the instances contained in the bag is positive and if a bag is labeled negative then every instance in the bag is negative. In our scenario, a bag represents a slide, and the instances in the bag are the images of the slide obtained by the MCAS. The slides are labeled as either adenocarcinoma positive or negative.

To test slide-level cancer detection with whole cytology smears, we used 16 labeled slides: eight that contained adenocarcinoma (positive) and eight that did not (negative). To address this limited number of available slides, we employed two strategies. First, we augmented the dataset by creating multiple bags from each slide (rather than using one bag per slide). Since the cellular material is distributed over a large area on a cytology smear, we assume that for a positive slide, the image obtained by each MCAS sensor (3072$\times$3072 pixels) contains a region that exhibits features associated with adenocarcinoma (Fig.~\ref{fig:slide_level_classification}b). So, we created bags of nine images each, by dividing sensor images into nine sub-patches of 1024$\times$1024 pixels, and eight such bags were created per slide amounting to 128 total bags. All the bags from a positive slide retained the positive label as per our assumption above, and all bags from negative slides retained a negative label as there would be no features associated with adenocarcinoma anywhere on those slides (Fig.~\ref{fig:slide_level_classification}c).

Then, we conducted a 4-fold cross-validation, which helped us get a better estimate of the algorithm performance with limited data. We divided the data into four groups of four slides each, where each group contained two positive slides and two negative slides. All the bags created from a given slide remained in a single group. To perform the cross-validation, three groups (12 slides, six positive and six negative, amounting to 96 bags) were used to train the model, and the remaining group (four separate slides, two positive and two negative, amounting to 32 bags) was used to test the trained model. This process was repeated four times, such that each group served as the test set in one of the iterations. The mean performance across the four iterations is used to evaluate success. At the bag level, our results demonstrated an average test accuracy of 0.930, precision of 0.931, recall of 0.938, and AUC of 0.969 (Fig.~\ref{fig:slide_level_classification}d). We also obtained a slide-level label by aggregating the bag-level outcomes via majority voting. This resulted in a slide-level test accuracy of 100\% across all 4 folds. An overview of the MIL model training is provided in Supplementary Fig. 2.

\section{Discussion}
We developed the MCAS to parallelize microscopic imaging of cytopathology slides, enabling significantly faster 3D digitization of this sample category (several minutes vs. hours). The MCAS can digitize slides using 0.3 and 0.5 NA lenses for visual inspection by pathologists, offering a similar price point for required components but with orders of magnitude faster 3D slide scanning compared to current WSI systems. In the future, mass production of MCAS lenses, the use of less expensive electronic components, and less precise translation stages could drive the cost down even further. Moreover, we employed machine learning algorithms to process these large image datasets from 0.3 NA lenses for adequacy assessment, cancer localization, and slide classification.

To further increase data acquisition speed, we can consider two possible approaches. First, we can reduce the number of scans in lateral directions. A 7x7 scanning grid was utilized to allow overlapping between FOVs for stitching. However, this overlap is not necessary for machine learning tasks that use image patches as inputs. By adopting a 6x6 scanning grid instead, we can reduce the acquisition time by 26 \%. Additionally, another way to increase speed is by increasing the axial step size required for capturing thick specimens. For a 0.3 NA lens, we used a \SI{5}{\micro\metre} step size to match its depth of field. While humans require crisp images and perfect focus, we hypothesize that machine learning algorithms can be trained using less than optimal focus images. Since the data transfer rate is the bottleneck for our data acquisition, doubling the step size (or skipping every other plane) could halve the acquisition time, assuming the axial scan range remains fixed. By combining these two strategies, we can reduce the total acquisition time by 63\%, and plan to demonstrate this in future work.

In the cancer localization task, we employed an object detection algorithm to identify key areas on the slide needing further examination by a pathologist. This approach aims to streamline the workflow for digital analysis of thick cytology smears. Inconsistent labeling, especially for adenocarcinoma at the cellular level, is a major challenge that affects network performance. While clinical practice does not require identifying every adenocarcinoma instance, accurately labeling all positive areas is essential for effective object detection. The high volume of data requiring annotation often leads to discrepancies in bounding boxes, hindering model training and producing suboptimal results. However, our results demonstrate the potential of the MCAS for rapid, automated cancer localization tasks with FNA smears. To mitigate the sparsely annotated dataset, we rescaled the loss for adenocarcinoma-positive labels to minimize false negatives. Our aim was to achieve higher recall rather than precision.

We demonstrate the effectiveness of MIL in slide-level disease identification. While this paper focuses on only two categories for the classification task, our approach can readily be applied to multi-class classification tasks or grading of tumors or disease severity in the future~\cite{pathak2014fully}. Additionally, our demonstration was based on the assumption that in adenocarcinoma-positive cases, each single FOV (3072$\times$3072 pixels) contains areas indicative of adenocarcinoma. If this assumption proves accurate, it could eliminate the need for scanning entire slides; instead, selecting a limited number of FOVs might be sufficient for diagnosis, further reducing the required scan time for future acquisition methods. With an MCAS and a trained network, scanning could cease as soon as one or enough patches indicative of cancer are detected, potentially leading to faster diagnoses.

Additional possible directions to improve MCAS digitization and automated analysis include enhancing its illumination and utilizing 3D data. This work used standard bright-field transmission illumination. Variable-angle illumination methods, for example as provided by a programmable LED array, may provide added morphological measurements to assist with localization and classification tasks. A learned sensing network~\cite{chaware2020towards,kim2020multi} could be adopted in the future to help establish optimal illumination strategies to maximize machine learning network performance. Additionally, we can use the entire 3D dataset instead of all-in-focus images for machine learning demonstrations. We anticipate that incorporating 3D data could further enhance model performance, which we plan to explore in future work.

Finally, we are planning to incorporate fluorescence capability into the MCAS. Fluorescence imaging is a versatile and non-invasive technique that enhances diagnostic accuracy, guides treatment decisions, and contributes to a deeper understanding of various diseases and physiological processes in clinical settings~\cite{conklin2009fluorescence}. We believe that fluorescence imaging offers numerous benefits in various fields, and our MCAS can expedite the imaging process, maximizing its advantages. In the future, such fluorescence imaging with the MCAS system could enable high throughput generation of image pairs to even digitize the staining process~\cite{kreiss2023digital}.

\section{Methods}
\subsection{MCAS data acquisition}
\begin{figure*}[ht]
    \centering
    \includegraphics[scale = 0.2]{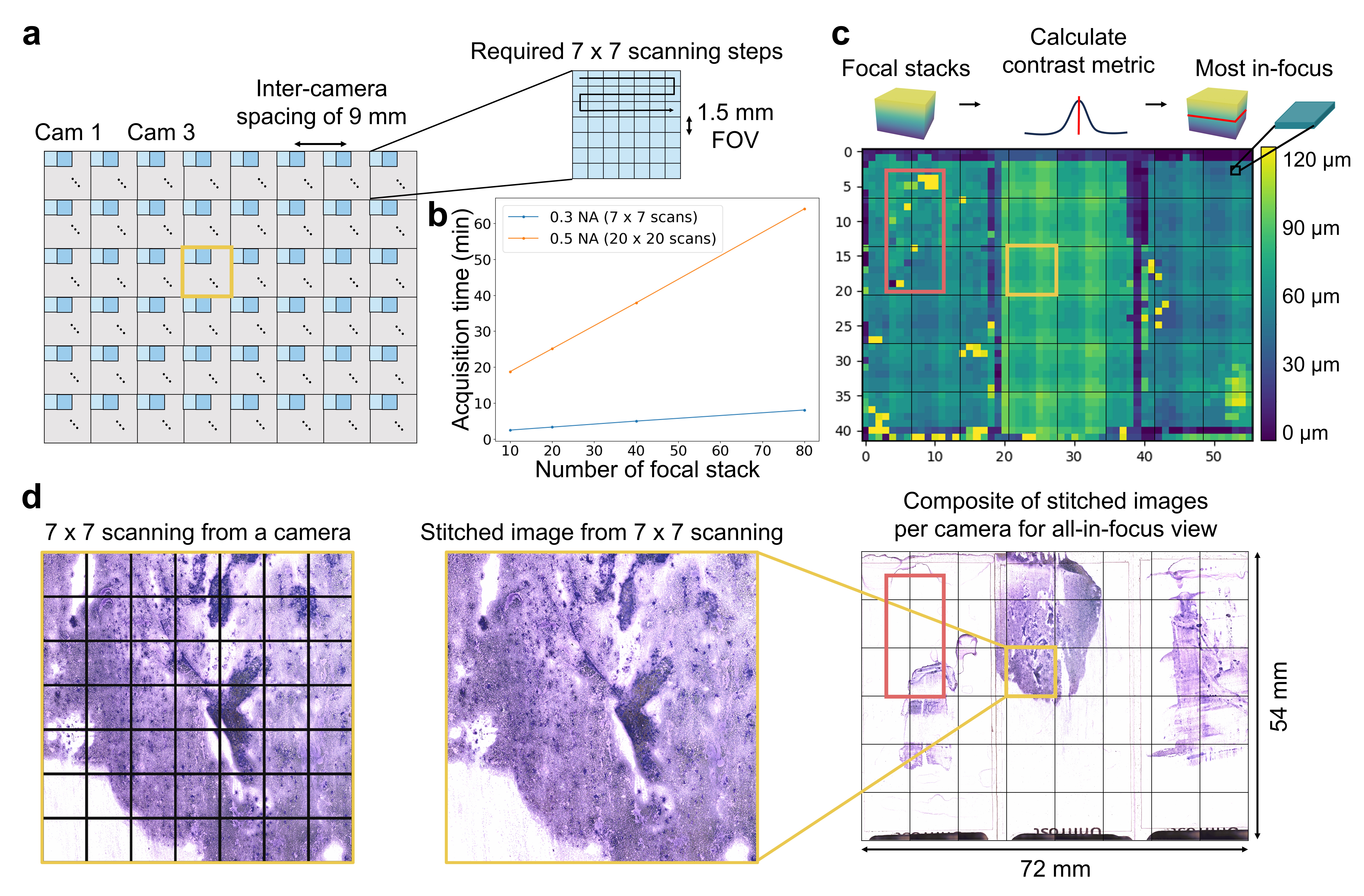}
    \caption{\textbf{The MCAS pipeline overview.} \textbf{a}-\textbf{b} The MCAS data acquisition (\textbf{a}) 48 cameras capture unique FOVs. A 7$\times$7 lateral scan is performed to fill FOV gaps with a 0.3 NA lens. (\textbf{b}) Axial scanning addresses the volume of thick samples. The plot shows the 3D data acquisition time per lens and a number of focal stacks. \textbf{c} Identifies the sharpest focus plane in each stack using a contrast metric. \textbf{d} In-focus planes from \textbf{c} are stitched together: first, 7$\times$7 scanned images from each camera, utilizing a 0.3 NA lens, are merged, then combined into a final composite after appropriate padding. Red boxes in \textbf{c} and \textbf{d} indicate FOVs where the contrast measurement failed due to the absence of distinctive sample features. The MCAS pipeline is consistent, differing only in the number of scans when utilizing a 0.5 NA lens.}
    \label{fig:workflow}
\end{figure*}

As discussed previously, the MCAS system requires lateral scanning to acquire images over a continuous FOV as the system’s magnification is greater than 1 ($M=2.2$ or $M=6.89$), the FOV of each lens-sensor pair (1.5 mm or 0.5 mm) is smaller than the camera sensor size (3 mm). Therefore, mechanical scanning is required to fill the gap between adjacent FOVs~\cite{harfouche2023imaging}. With 0.3 NA lens micro-cameras having a 1.5 mm FOV and 9 mm spacing between them, a grid format of at least 6$\times$6 is needed for a seamless image. In practice, we scan a 7$\times$7 grid with $\approx$1.3 mm step sizes to ensure that we have an overlap between adjacent FOVs to stitch images into a final composite (Fig.~\ref{fig:workflow}a). Lateral scanning increases with the magnification of the lens, necessitating a 20$\times$20 grid scan with our 0.5 NA lens with 0.46 mm step sizes (Supplementary Fig. 3). While existing sample-scanning approaches require the translation stages to traverse the entire desired FOV (in this case 54$\times$72 mm$^2$), our approach only requires translating up to the inter-camera spacing of 9$\times$9 mm$^2$ and hence is much faster. Also, it does not require scanning mechanics that must cover a long scan distance, allowing us to use more cost-effective and simpler stages. In addition to lateral scanning, we use axial scanning to capture 3D information from the thick cytology specimens. For the cytology slides used in this study, the scanning range spanned \SI[parse-numbers=false]{120-150}{\micro\metre} and we acquired data with either a \SI{5}{\micro\metre} or \SI{1}{\micro\metre} step increment, to match the lenses' depth of field of 0.3 or 0.5 NA, respectively. Arbitrary step sizes can be selected based on imaging requirements, and precise step increments can be achieved with an encoder.

The MCAS utilized 4:3 format CMOS sensors and we cropped and transferred only the center FOV (3072$\times$3072 pixels) to optimize data transfer speed. With this configuration, the system generated $\approx$0.45 GP per snapshot or $\approx$16 GP per focal plane which consists of 2352 (\# of camera$\times$\# of scanning = 42$\times$56) FOVs. With 0.5 NA lenses, the size of each final composite focal plane is 140 GP, consisting of 19200 FOVs. To achieve rapid data capture, we developed an `on-the-fly' method that captures and saves data simultaneously while the stages move axially. We established an optimal exposure time of 50 ms and a scan speed of \SI{34}{\micro\metre/\s} and \SI{7}{\micro\metre/\s} respectively to achieve this. This approach increased speed sevenfold compared to a stop-and-go capture method, where stages move to specific locations, capture data, and save it to storage step by step. The current data rate limit of 5 GP/second restricts faster axial scan speeds, with the potential for further speed improvements through system architecture enhancements. Data acquisition times for each lens type are shown in Fig.~\ref{fig:workflow}b. The time consumed by the lateral movement, assuming no scanning in the axial direction, is approximately 40 seconds for the 0.3 NA lens and 320 seconds for the 0.5 NA lens.

\subsection{MCAS image preprocessing}
After collecting 3D data from thick cytology specimens, we extract the most in-focus plane and adjacent planes above and below using a contrast metric based on the Laplacian with a 7x7 kernel size for each lateral scan position. These planes are displayed to the user, allowing for image focus adjustments during the inspection. Fig.~\ref{fig:workflow}c shows the axial heights of each micro-image in the composite. Height measurements may fail due to the contrast measurement's sensitivity to sensor noise or the presence of dust on the cover slip. The red boxes in Fig.~\ref{fig:workflow}c,d indicate where the contrast measurement failed. Additionally, there is a significant height change between the top three rows due to the cover slip.

Subsequently, we conducted a white balance procedure to ensure accurate color representation in images examined by pathologists for diagnostic purposes. White balance involved adjusting the red, green, and blue channels to portray colors in the sample accurately. This step is crucial as the appearance of images can vary based on light sources and sensor types. To achieve white balance, we performed histogram matching between acquired and reference images for each RGB channel. The reference images were obtained from clinical settings to ensure that the balanced images closely resemble those seen by pathologists. Additionally, we applied a flat-field correction to compensate for lens vignetting and illumination brightness variations before stitching images.

\subsection{MCAS image stitching}
To stitch the processed images, we utilized the open-source Hugin code base~\cite{hugin}. This stitching process involves global transformations that apply lateral shifts, rotational corrections, and scaling corrections to match the images. Since the stitching algorithm relies on features across multiple images, areas without specimen material present a challenge. We addressed this by generating stitching parameters from a printed target, which were then used to stitch our slide images accurately. Stitching 49 images (from 49 lateral scanning positions) from a camera takes less than a minute and an additional 9-10 minutes to correct image intensity variation. This stitching process can be parallelized as images from each camera can be processed independently. See Supplementary Fig. 4 for additional details. Stitching is not essential for disease detection and classification, as a specific model can be applied directly to raw image data. However, for tasks like adequacy assessment, stitching is a beneficial post-processing step.

\subsection{MCAS visualization}
Effective visualization of MCAS images presents two main challenges. First, the final all-in-focus image generated from the data processing steps described previously is 16 GP, which is more than can be displayed simultaneously on a single monitor. Additionally, sometimes the height variation within a single camera FOV exceeds the lens' depth of field, so we cannot guarantee that every point within the sensor image is in focus.

To address challenges for seamless viewing of high-resolution images, we developed Gigaviewer, a custom-built, web-based viewer using OpenSeadragon~\cite{openseadragon}. It includes user-adjustable focus capability on top of zoom and pan options provided by OpenSeadragon. The viewer generates images at different zoom levels to show varying resolutions, preloading adjacent fields of view for seamless viewing. Example images can be viewed at \url{https://gigazoom.rc.duke.edu}.

Adjusting the focal plane, as pathologists do with traditional microscopes, is crucial for examining thick specimens with varying heights. To ensure every part of the specimen is in focus, we stitched and visualized images for the most in-focus plane as determined by the Laplacian metric, as well as the planes above and below it. This allows users to switch between focal planes and view thick specimens without defocusing problems.

\subsection{MCAS hardware}
For illumination, we utilized a 16$\times$24 LED array with a 9 mm LED pitch (384-well RGB LED microplate light panels from Tindie), positioned approximately 5 cm beneath the sample plane. A diffusive screen was placed above the LED array. To scan specimens laterally and axially, we employed three motorized linear stages from Zaber (X-LSM Series with travel distances 200, 100, and 50 mm). Table~\ref{tab:Lens_specifications} lists the main optical characteristics of the customized lenses used in our MCAS, and Supplementary Fig. 5 shows the characterization of resolution uniformity.

\begin{table}[h]
\addtolength{\tabcolsep}{0pt}
  \caption{\bf Lens specifications}
  \label{tab:Lens_specifications}
  \centering
  \begin{tabular}{lcc}
    \toprule
    Characteristic & \multicolumn{2}{c}{Specification}\\
    \cmidrule{2-3}
    & 0.3 NA & 0.5 NA\\
    \midrule
     Lens diameter & 8 mm & 8.5 mm\\
     Lens length & 26 mm & 28 mm\\
     Focal length & 4.95 mm & 6.56 mm\\
     Working distance &  3.05 mm & 1.744 mm\\
     Object space NA & 0.3 & 0.5\\
     Magnification & 2.2X & 6.89X\\
     Designed wavelength & 470-660 nm &  486-656 nm\\
     Primary wavelength & 520 nm & 587 nm\\
    \bottomrule
  \end{tabular}
\end{table}

As shown in the Supplementary Fig. 5, there is a one-element drop in resolution for both the 0.3 NA and 0.5 NA lenses when comparing the center and the edge of the FOV. Since our annotations are present in both the center and the edge of the FOV, our trained model accounts for all possible aberrations in our imaging system, enhancing its robustness. For visual smear inspection, where minimizing the effects of resolution degradation and lens aberration at the edge of the FOV is beneficial, we plan to crop out the outermost FOV before using it for future studies.

Also, this imaging system exhibits field curvature. To measure it, we imaged a flat target using axial scanning with a \SI{5}{\micro\metre} step size. We then identified the most in-focus plane at each pixel using contrast measurements obtained with a Laplacian operator~\cite{zhou2024computational}. Ideally, only one focal plane should be used across the FOV to form the all-in-focus image, as the object is flat. However, as shown in the Supplementary Fig. 6, our imaging system exhibits field curvature, with each lens showing a slightly different amount. Since we capture a focal stack to image thick specimens, We can partially correct for this field curvature by using such a map to adjust the z-slices that different pixels are assigned to. In the majority of this work, including for the developed machine learning applications, we generated all-in-focus images before feeding them to the networks, minimizing the effect of defocus.

\section*{Acknowledgments}
Research reported in this publication was supported by the National Cancer Institute (NCI) of the National Institutes of Health under award number R44CA250877, the Office of Research Infrastructure Programs (ORIP), Office of the Director, National Institutes of Health of the National Institutes of Health and the National Institute of Environmental Health Sciences (NIEHS) of the National Institutes of Health under award number R44OD024879, the National Institute of Biomedical Imaging and Bioengineering (NIBIB) of the National Institutes of Health under award number R43EB030979, the National Science Foundation under award number 2036439 and award number 2238845, and the Duke Coulter Translational Partnership Award. This project has received funding from the European Union’s Horizon 2022 Marie Skłodowska-Curie Action under grant agreement 101103200. 

\section*{Supplementary information}
See Supplementary document for supporting content.

\section*{Competing interests}
R.H. and M.H. are cofounders of Ramona Optics, Inc., which is commercializing multi-camera array microscopes. C.B.C., M.H., P.R., V.S., J.D., C.D., and G.H. are or were employed by Ramona Optics, Inc. during the course of this research. K.C.Z. was a consultant for Ramona Optics, Inc. The remaining authors declare no competing interests.

\section*{Data availability}
Processing code and data can be found at (\url{https://github.com/powergkrry/MCAS-Cytology}). Additional data underlying the results are available from the corresponding authors upon reasonable request.

\section*{Authors' contributions}
K.K. and R.H. conceived the idea and initiated the research. With the help of A.C., C.B.C., S.X., K.C.Z. and R.H., K.K. developed the algorithms and theory. K.K. and A.C. wrote the code for and performed data analysis. K.K. and C.C. developed customized Gigaviewer. K.K., M.H., P.R., V.S., J.D., C.D., G.H., and R.H. developed the MCAM hardware and acquisition software. With the help of M.A., R.D., I.T.C., W.C.F., X.S.J., and R.H., K.K. acquired and analyzed the cytology smear data. M.A., R.D., I.T.C., W.C.F., and X.S.J. provided input to and performed data annotation. K.K. and A.C. wrote the manuscript and created the figures, with input from all authors. K.K, A.C., L.K. and R.H. revised the manuscript. R.H. supervised the research.

\bibliographystyle{unsrt}  
\bibliography{references}

\appendix
\section{Supplementary Figure S1 Resolution analysis}

\begin{figure}[h]
    \centering
    \includegraphics[scale = 0.2]{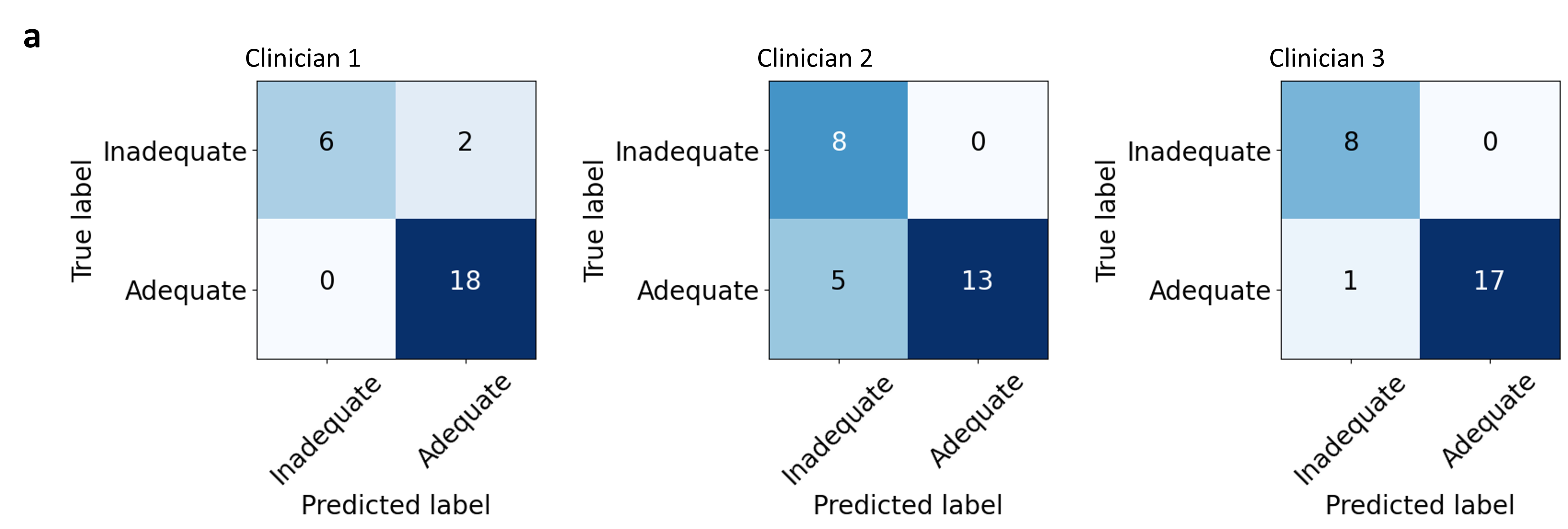}
    \caption{\textbf{Evaluation of thyroid specimen adequacy.} \textbf{a} Detailed results from each clinician's assessment}
    \label{fig:ROSE_each}
\end{figure}

\begin{figure}[h]
    \centering
    \includegraphics[scale = 0.2]{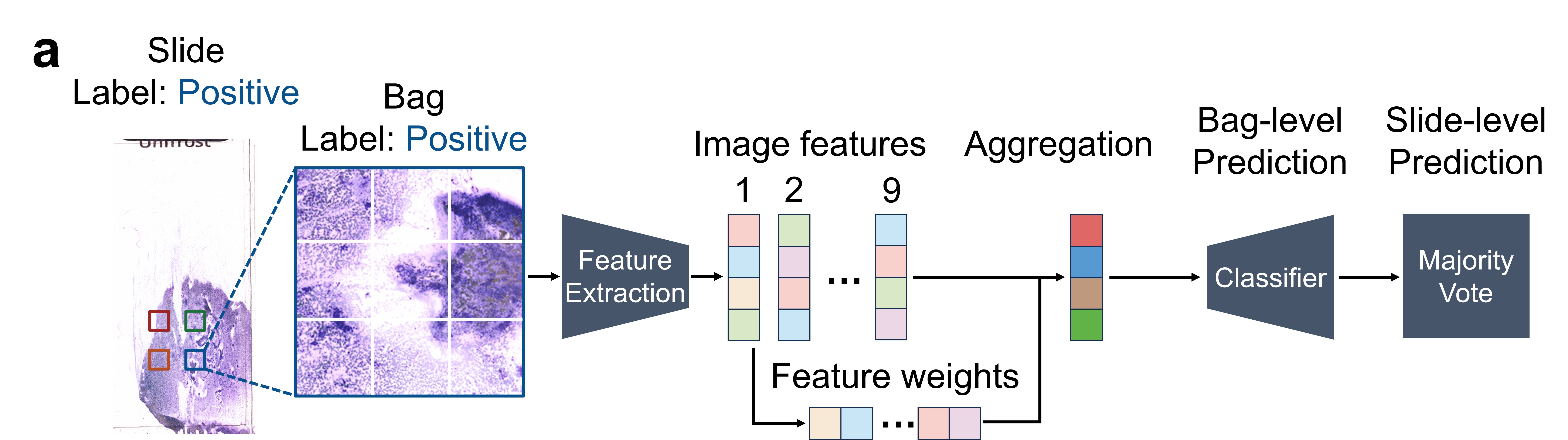}
    \caption{\textbf{Overview of Multiple Instance Learning.} \textbf{a} We have slide-level labels for each slide, and we use the MCAS system to image these slides. For each field of view, we generate nine `instances' to create a `bag'. All bags derived from a single slide are assigned the same label. Then, we extract image features using a feature extraction layer and feed these features into an attention layer, which generates feature weights. This layer assigns weights to each image feature, and the weights are used to compute a weighted sum of the image features. The aggregated feature is then passed into a classifier layer to predict the label for the bag. For slide-level label prediction, a majority vote among all bags from a given slide is used.}
    \label{fig:MIL_slidelevel}
\end{figure}

\begin{figure}[h]
    \centering
    \includegraphics[scale = 0.2]{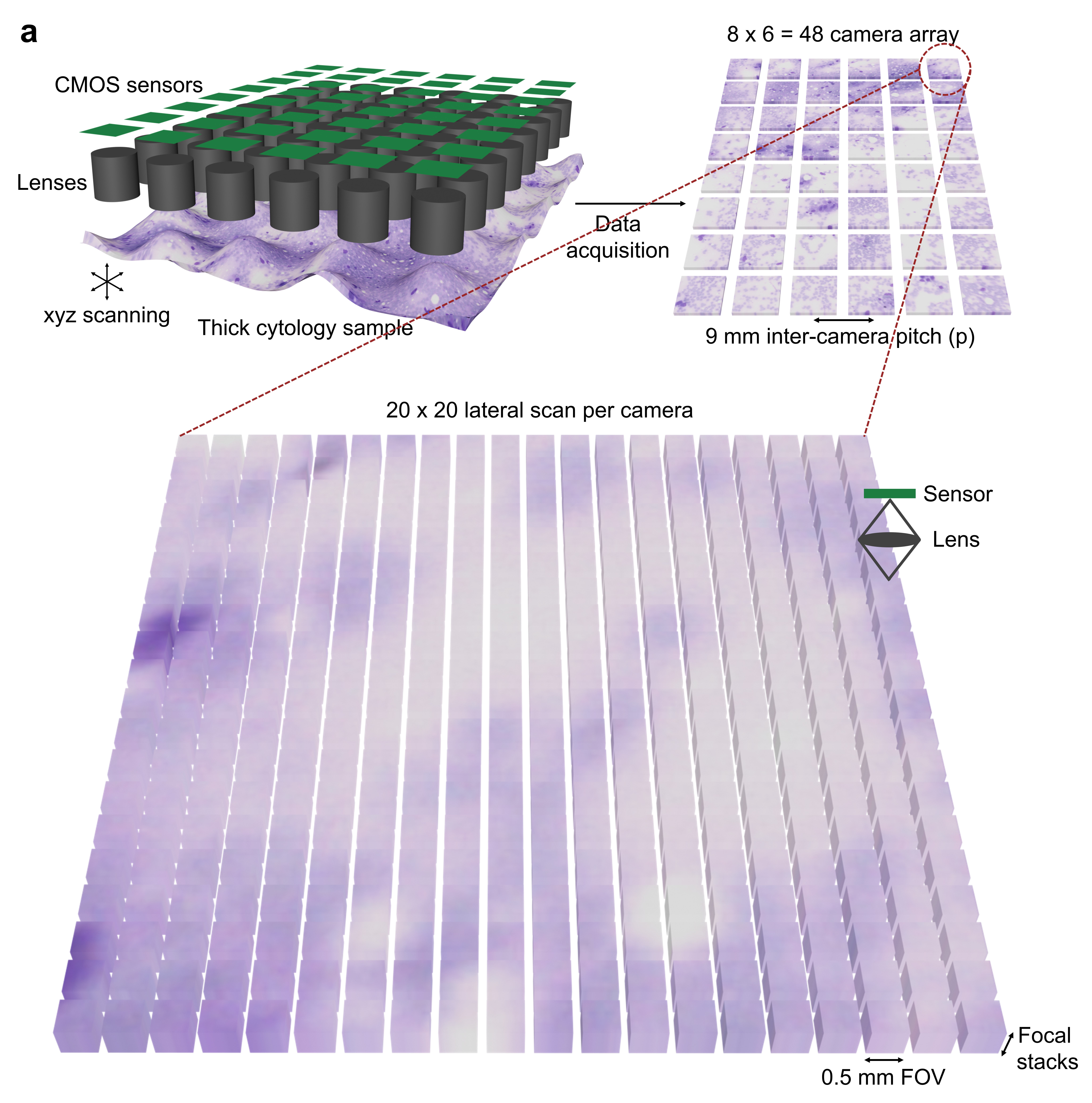}
    \caption{\textbf{Data acquisition using the Multi-Camera Array Scanner (MCAS) equipped with 0.5 NA lenses.} \textbf{a} Rapid 3D imaging is achieved with 48 cameras, each offering a unique 0.5 mm field of view. A 20×20 lateral scanning grid compensates for the 9 mm inter-camera gaps, and axial scanning is utilized to capture surface irregularities in thick specimens.}
    \label{fig:20x_diagram}
\end{figure}

\begin{figure}[h]
    \centering
    \includegraphics[scale = 0.175]{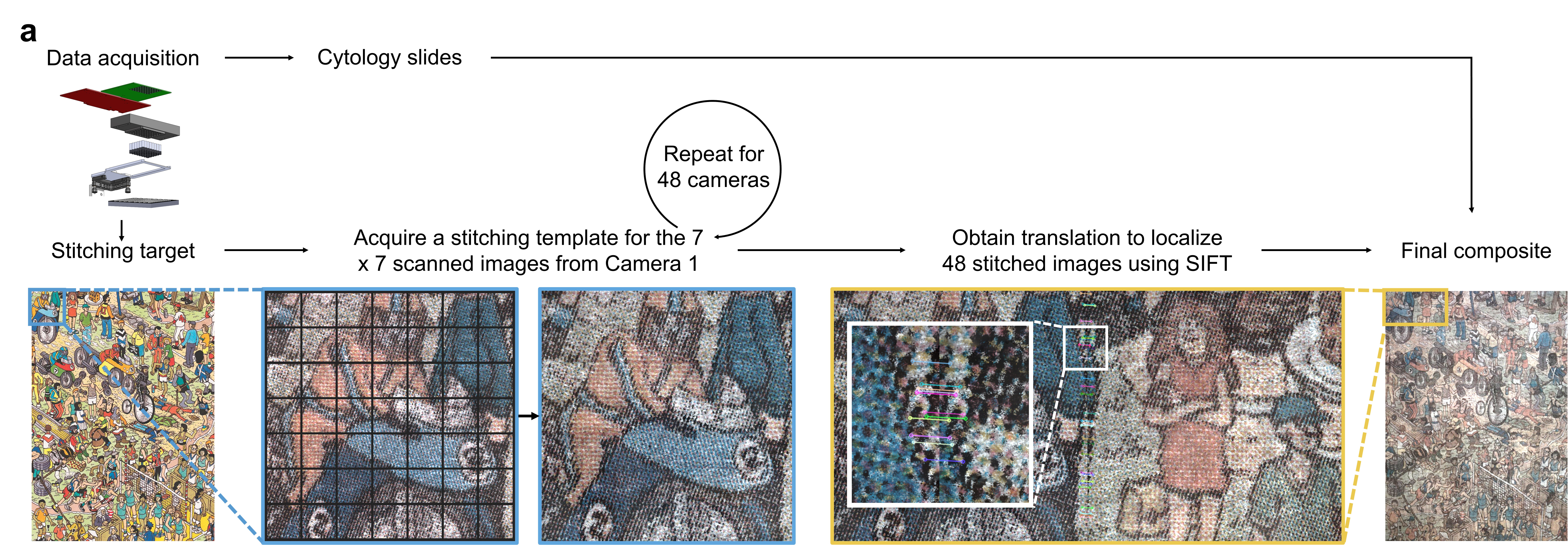}
    \caption{\textbf{Stitching workflow using calibration target.} \textbf{a} To stitch the processed images, we first obtained calibration stitching parameters using a flat target. We acquired two sets of parameters: one for images from each camera and one for the final composite. Initially, we found parameters for images from a single camera's 7$\times$7 scanning positions and repeated this for 6x8 cameras individually. Following that, we obtained parameters for the final composite using scale-invariant feature transform (SIFT). This method extracted features and calculated the average x-y shift in the features from abutting images. With these parameters in hand, we stitched images from each camera separately and padded the stitched images to their appropriate location. It is important to note that in the second step, we were unable to utilize the Hugin stitching process due to the immense size of the total image. As a result, we only considered translation for localization. The final composite is the output utilizing the most in-focus plane, referred to as an `all-in-focus' image (see Fig. 6d).}
    \label{fig:image_stitching}
\end{figure}

\begin{figure}[h]
    \centering
    \includegraphics[scale = 0.45]{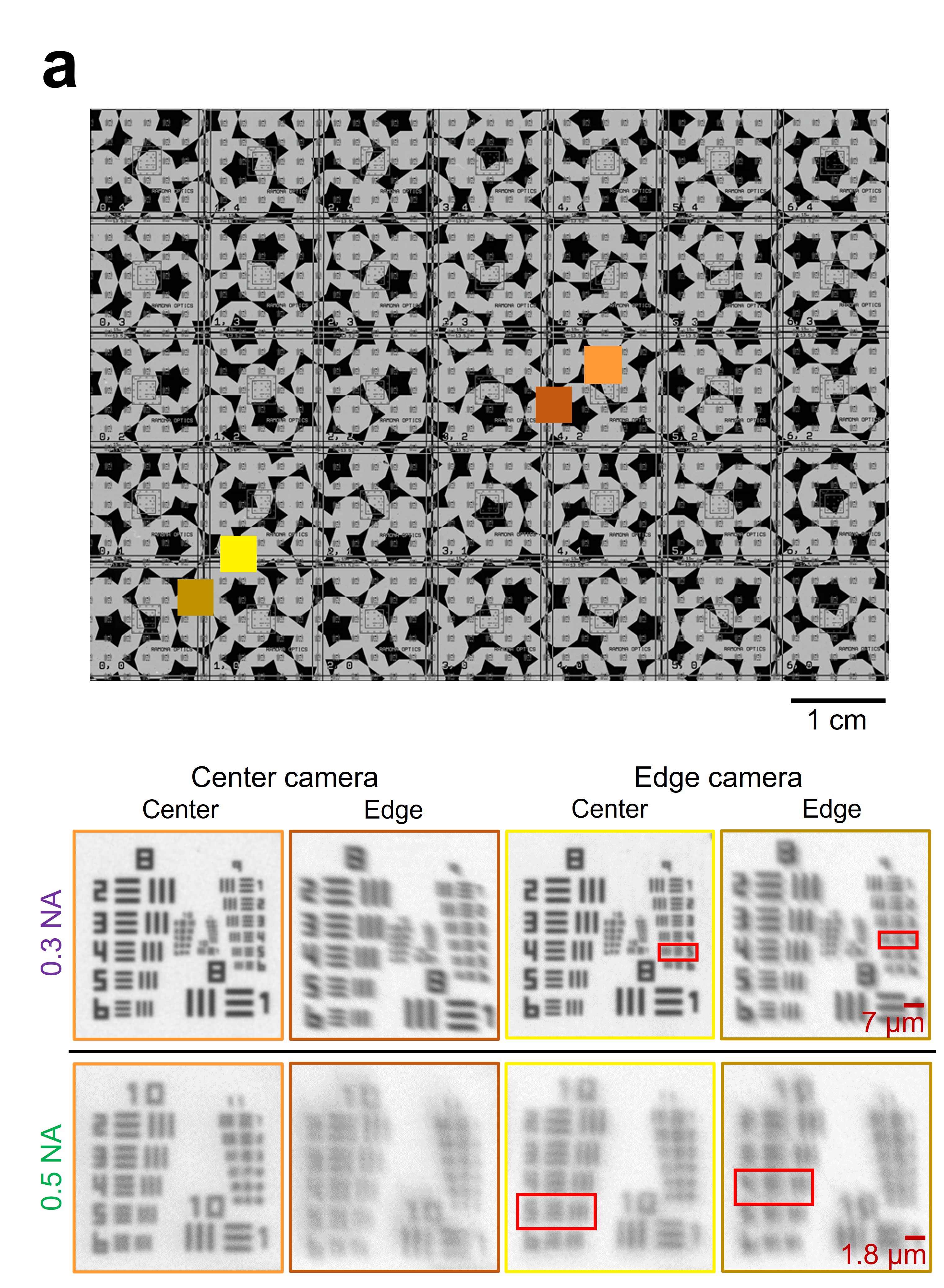}
    \caption{\textbf{MCAS resolution assessment across different cameras
and across the entire FOV.} \textbf{a} The top row represents the 0.3 NA lens, while the bottom row represents the 0.5 NA lens. Two lenses of each type were used for this analysis, and the central and peripheral resolutions for each lens are reported.
}
    \label{fig:resolution_assessment}
\end{figure}

\begin{figure}[h]
    \centering
    \includegraphics[scale = 0.4]{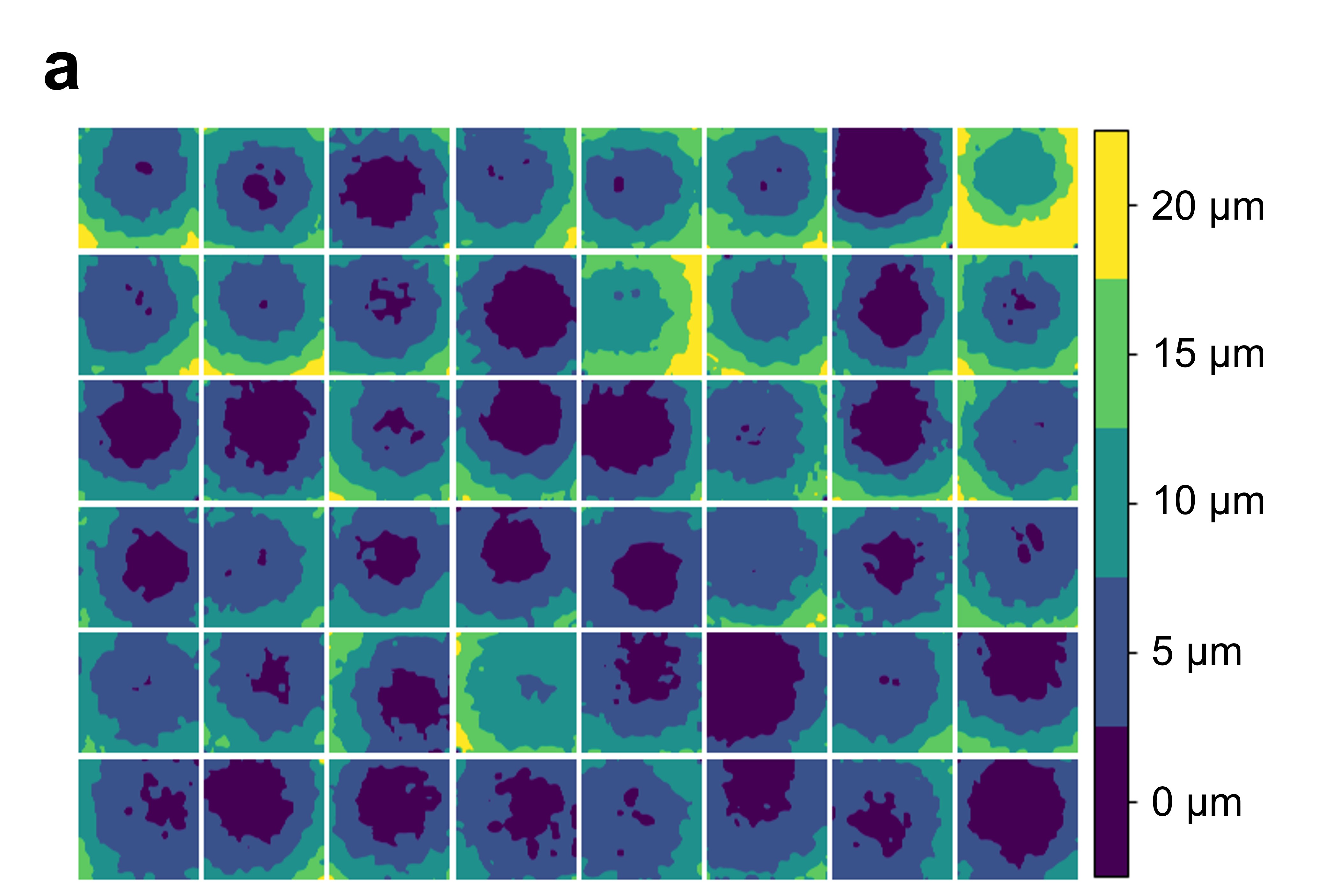}
    \caption{\textbf{Field curvature.} \textbf{a} The field curvature for 48 lenses is reported.}
    \label{fig:field_curvature}
\end{figure}

\end{document}